# Experimental detection of topological electronic state and large linear magnetoresistance in SrSn$_4$ superconductor

Arnab Kumar Pariari[1,2,*], Rajesh O Sharma[1], Mohammad Balal[3], Markus Hücker[2], Tanmoy Das[1], Sudipta Roy Barman[3]

[1] Department of Physics, Indian Institute of Science, CV Raman Rd, Bengaluru, Karnataka 560012, India

[2] Department of Condensed Matter Physics, Weizmann Institute of Science, 234 Herzl Street, POB 26, Rehovot 7610001 Israel

[3] UGC-DAE Consortium for Scientific Research, Khandwa Road, Indore 452001, Madhya Pradesh, India

[*] Corresponding author: arnabpariari@iisc.ac.in, pariari.arnab@gmail.com

While recent experiments confirm the existence of hundreds of topological electronic materials, only a few exhibit the coexistence of superconductivity and a topological electronic state. These compounds attract significant attention in forefront research because of the potential for the existence of topological superconductivity, paving the way for future technological advancements. SrSn$_4$ is known for exhibiting unusual superconductivity below the transition temperature (T$_C$) of 4.8 K. Recent theory predicts a topological electronic state in this compound, which is yet to be confirmed by experiments. Systematic and detailed studies of the magnetotransport properties of SrSn4 and its Fermi surface characterizations are also absent. For the first time, a quantum oscillation study reveals a nontrivial $\pi$-Berry phase, very light effective mass, and high quantum mobility of charge carriers in SrSn$_4$. Magnetotransport experiment unveils large linear transverse magnetoresistance (TMR) of more than 1200% at 5 K and 14 T. Angle-dependent transport experiments detect anisotropic and four-fold symmetric TMR, with the maximum value (~ 2000%) occurring when the angle between the magnetic field and the crystallographic b-axis is 45$^0$. Our results suggest that SrSn$_4$ is the first topological material with superconductivity above the boiling point of helium that displays such high magnetoresistance.

**Keywords**

Topological electronic materials, Superconductivity, Electrical transport, Magnetism, Quantum Oscillations, Fermi Surface

## 1. Introduction

New materials prediction and experimental discovery are at the forefront of current condensed matter research, particularly in the field of topological quantum materials.[1-7] These materials manifest intriguing quantum phenomena of fundamental importance while many of their unique electronic properties are also crucial for technological advancements. These properties include very large magnetoresistance, high charge carrier mobility, anisotropic magnetoresistance, intrinsic anomalous Hall effect, large thermoelectric power, anomalous Nernst effect, unconventional superconductivity, and helicity-dependent photocurrents.[8-20] Compounds that display multiple such electronic properties garner special attention from the scientific community due to their potential for versatile applications. The rare coexistence of topological electronic states and superconductivity holds promise for uncovering unconventional superconductivity, particularly spin triplet superconductivity, which could be important for quantum computing applications.[21] The compound of this study, SrSn$_4$, has a superconducting transition temperature of $T_c = 4.8$ K, and possesses features that deviate from the behavior anticipated for an isotropic, single-band weakly coupled Bardeen-Cooper-Schrieffer (BCS) superconductor.[22,23] It has been proposed that this unusual superconductivity in SrSn$_4$ is because of multi-band superconductivity.[23] Recent theoretical databases of proposed topological materials identify SrSn$_4$ as a candidate along with thousands of other compounds.[1-3] In particular, SrSn4 is listed as a topological semimetal when SOC is excluded, and as a three-dimensional (3D) topological insulator (TI) when SOC is included. A stoichiometric topological compound with such a high $T_c$ at ambient pressure is extremely rare in the literature. There are only very few experimentally verified topological materials, such as MgB$_2$, NaAlSi, and β-PdBi$_2$ that have a higher $T_c$ than SrSn$_4$.[24-27] However, there are no experimental studies on SrSn$_4$ so far on topological electronic state, detailed magnetotransport properties, and characterizations of the Fermi surface. Here we report the first observation of quantum oscillations in SrSn$_4$ in both magnetic and transport measurements. Our analysis of the quantum oscillations provides clear evidence of a topological electronic state. Magnetotransport experiments reveal a very large, nonsaturating, linear, and anisotropic transverse magnetoresistance (TMR). Although there have been reports on the coexistence of superconductivity and topological electronic states in several compounds, only a few of them are known to exhibit large magnetoresistance, such as PbTaSe$_2$, Au$_2$Pb, YPtBi, PbTe$_2$, MoTe$_2$, α-Ga, γ-PtBi$_2$, AuSn$_4$, PtPb$_4$, CaSb$_2$, and SnTaS$_2$.[28-43] Remarkably, the superconducting transition of all these compounds is significantly below the boiling point of helium, which is a critical threshold for low-temperature experiments and applications. Below this threshold, supplementary cooling techniques are utilized instead of solely relying on a liquid helium bath or a basic 4 K closed-cycle cryostat operating on the Gifford-McMahon refrigeration mechanism. The present study to the best of our knowledge identifies SrSn$_4$ as the first topological compound, exhibiting $T_c$ above the boiling point of helium at ambient pressure, while also demonstrating remarkably high magnetoresistance.

## 2. Results and Discussions

### 2.1. Crystal structure and characteristic superconductivity

Single crystals of SrSn$_4$ were grown by the Sn flux technique, as described in the experimental details. Single crystal x-ray diffraction (details in experimental section) of as-grown SrSn$_4$ crystals at 100 K reveals the orthorhombic structure with space group *Cmcm*, as shown in **Figure** 1a, consistent with previous reports.[22,23] For the refined lattice parameters at 100 K we obtained $a = 4.5812(8)$ Å, $b = 17.315(4)$ Å, $c = 7.0242(13)$ Å, and $\alpha = 90^0$, $\beta = 90^0$, $\gamma = 90^0$, where $a$, $b$, and $c$ are naturally slightly smaller than the room temperature values reported in literature.[22,23] The crystallographic axes are identified as illustrated in the bottom right inset of **Figure 1**(b).

The superconducting state of SrSn$_4$ was characterized by measuring the temperature dependence of resistivity ($\rho_{xx}$) and magnetization (*M*) under varying external magnetic fields (H). **Figure 1**b shows a highly metallic resistivity with a large residual resistivity ratio (RRR) of 140, calculated as the ratio of $\rho_{xx}$ at 300 K to $\rho_{xx}$ at 5 K, followed by a superconducting transition with an onset temperature of $T_c$ = 4.8 K. Below $T_c$ the resistivity shows a sharp drop, reaching the resolution limit of our transport setup, and remains constant down to 1.8 K. Our $\rho_{xx}$ (T) data is in good agreement with the literature results for $T_c$ in SrSn$_4$.[22,23] If we compare $\rho_{xx}$ (T) in the normal state with literature, we find that the residual resistivity ratio (RRR) of the present sample is significantly larger, while the room temperature resistivity is substantially lower than 200 μΩ cm.[23] This could be an indication of a lower impurity and defect concentration in our crystal. $\rho_{xx}$ (T) under varying external magnetic field, as shown in **Figure 1**c, displays a large upper critical field H$_{c2}$ ~ 1 T at 2 K, which is much higher than a previously reported value of ~ 0.2 T.[23] Studies have shown that moderate impurity concentrations can lead to an enhancement of H$_{c2}$ in a superconductor without significantly affecting its $T_c$.[44-46] However, the presence of profound quantum oscillations and high quantum mobility of charge carriers (see next), very low resistivity, and larger RRR indicate that such a higher H$_{c2}$ might not simply be attributed to a higher degree of impurity in the present sample. Future research on superconductivity in SrSn$_4$ will help address this anomaly. In the zero-field-cooled (ZFC) dc magnetic susceptibility (*M/H*) shown in **Figure 1**d, superconductivity sets in at approximately 4.6 K. This value of $T_c$ onset from magnetization measurements is in agreement with the previously reported value.[23] *M/H* at some higher representative magnetic fields are shown in Figure S1b, Supporting Information. The strong Meissner effect, even at a 0.2 T magnetic field, indicates that H$_{c2}$ is higher than the previously reported value, which is consistent with our finding from the transport experiment.

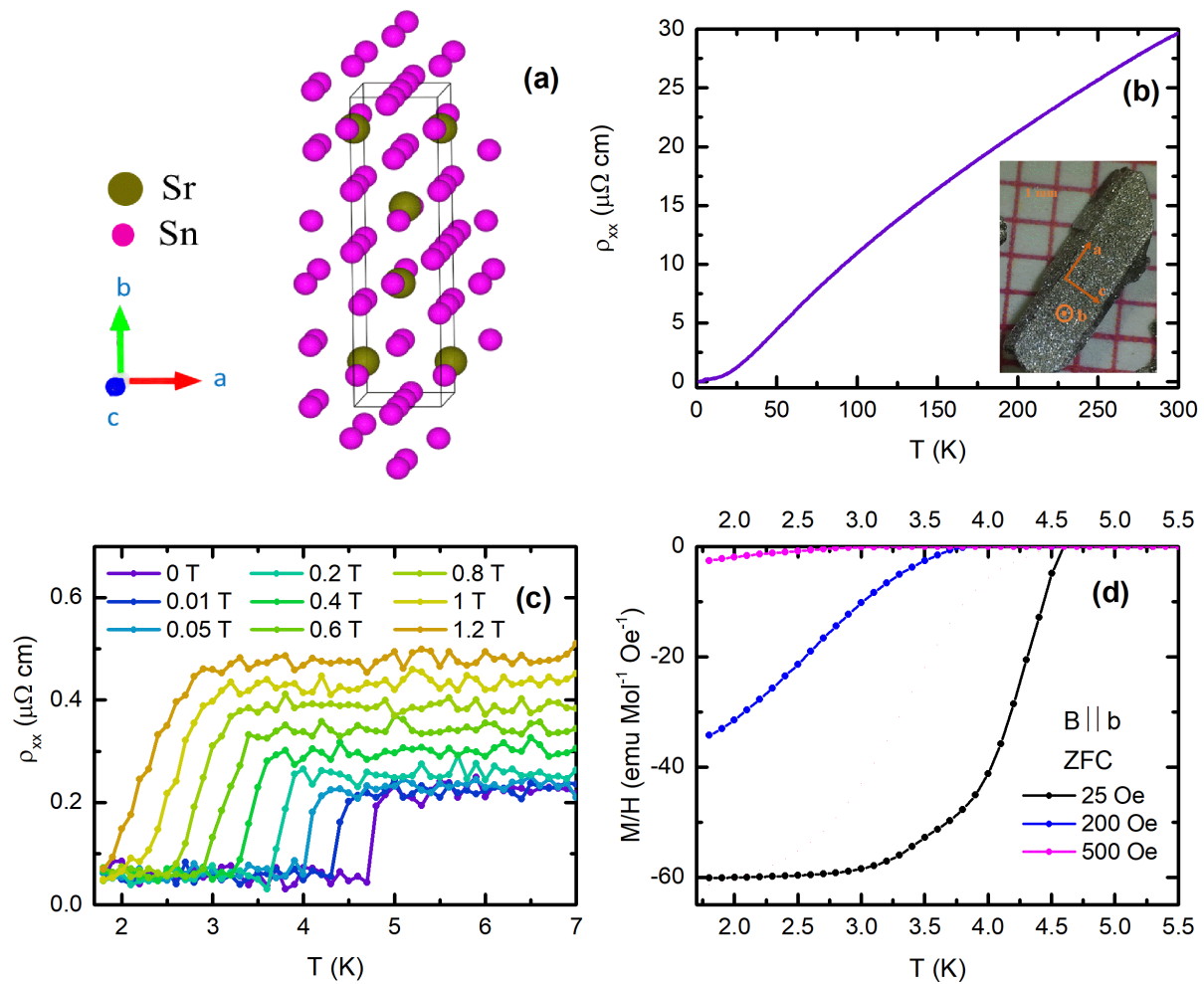

**Figure 1**: Structural characterizations and characteristic superconductivity of SrSn$_4$ single crystals. a) Crystal structure image of SrSn$_4$. It is generated using the crystallographic information file (see additional files) of the compound obtained in single crystal x-ray diffraction experiments and analysis. b) Temperature (*T*) dependence of resistivity ($\rho_{xx}$) from 1.8 K to

300 K. The bottom right inset shows an image of a representative as-grown crystal on millimeter-sized checkered paper. The three perpendicular crystallographic directions of the crystal are labeled, as determined in single-crystal X-ray diffraction experiments. (c) $\rho_{xx}$ ($T$) across the superconducting transition temperature ($T_c$) under external magnetic fields. (d) Zero-field-cooled dc magnetic susceptibility ($M/H$, where $M$ is the magnetic moment and $H$ is the field) in the temperature range of 1.8 K to 5.5 K.

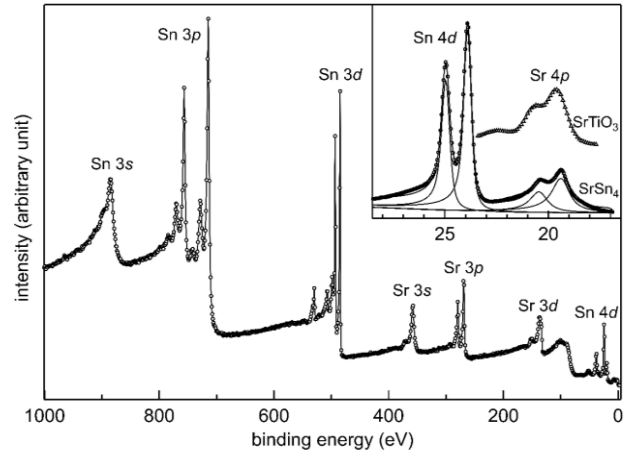

**Figure 2:** HAXPES survey spectra of SrSn$_4$ measured with 6 keV photon energy at room temperature. The main core-level peaks are labeled. The inset shows the Sn 4$d$ and Sr 4$p$ spectra (open circles) of SrSn$_4$ taken with a smaller step size along with the fitted curves (black). A comparison with the Sr 4$p$ spectrum of SrTiO$_3$ (open triangles, staggered vertically) is also shown.

Hard x-ray photoelectron spectroscopy (HAXPES) is often used to study the bulk electronic structure of materials to avoid the influence of surface effects.[47-49] For SrSn$_4$, the inelastic mean free path length of the photoelectrons using 6 keV photon energy is around 10 nm. A HAXPES survey spectrum of SrSn$_4$ is shown in **Figure 2**, where all the different core-level peaks of both Sn and Sr are observed. The inset shows an expanded view of the Sn 4$d$ and Sr 4$p$ spectra, which have been fitted using the nonlinear least square curve fitting method. We obtain the binding energy position of Sn 4$d$ in SrSn$_4$ to be 23.9 eV (24.95 eV) for the $4d_{5/2}$ ($4d_{3/2}$) spin-orbit component, which is close to that of Sn metal.[50] We find that the binding energy of Sr 4$p$ is 19.4 (20.5) eV for the $4p_{3/2}$ ($4p_{1/2}$) level. As shown by comparison in the inset, these are close to the HAXPES spectrum of SrTiO$_3$, where the Sr $4p_{3/2}$ peak appears at 19.6 eV. It is well known that Sr is in a +2 oxidation state in SrTiO$_3$, and the proximity of the binding energy of Sr 4$p$ of SrSn$_4$ to it shows that the latter is also in a +2 state.

## 2.2. de Haas-van Alphen (dHvA) oscillations and Fermi surface characterizations

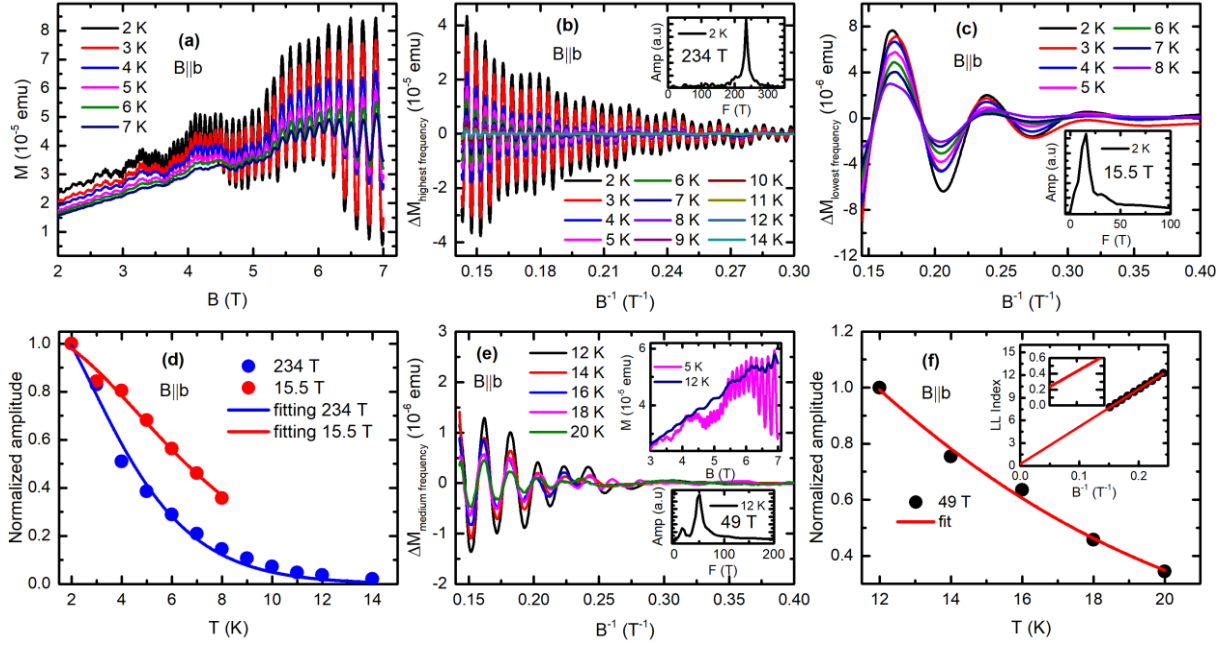

**Figure 2**: dHvA oscillations for magnetic field applied along *b*-axis direction in the SrSn$_4$ single crystals. a) *M* vs *B*, for magnetic field parallel to *b*-axis configuration. b) High frequency component of the oscillating part of the measured magnetic moment, *ΔM*, after background subtraction and deconvolution. The inset shows the peak position at 234 T in fast Fourier transformation (FFT) of the oscillations. c) The low frequency component of *ΔM*. Inset shows the corresponding FFT at 15.5 T. d) Temperature dependence of oscillation amplitudes (normalized) at a specific peak position of *ΔM* vs $\frac{1}{B}$ plots, for the two frequency components. The temperature dependence of oscillation amplitudes is fitted with the thermal damping term of the Lifshitz Kosevich (LK) formula. e) The oscillations in the low frequency component of *ΔM* which dominate over the 15.5 T frequency component at higher temperatures. The inset shows the FFT peak at 49 T. It is named as *ΔM*$_{\text{medium frequency}}$. f) Theoretical fit to the thermal damping of oscillation amplitude for the 49 T frequency. The Landau level index plot is shown in the inset. Following the convention, the valleys in *ΔM* are assigned as $n - \frac{1}{4}$ and peaks are assigned as $n + \frac{1}{4}$, where the Landau level index, *n* is an integer.

In several previous reports on topological compounds, it was found that de Haas-van Alphen (dHvA) oscillations in magnetization can be well detected in relatively low magnetic fields and moderately low temperatures, compared to Shubnikov-de Haas (SdH) oscillations in resistivity.[51-60] This is primarily because the SdH effect arises from oscillations in the scattering rate of charge carriers, unlike the dHvA effect, which originates from oscillations in free energy in the presence of a magnetic field. Being specific to the scattering rate, the SdH effect is influenced by the detailed scattering mechanism in a system. Also, it is affected by quantum interference as well as other electronic transport-related issues such as Joule heating and noise from electrical contacts. Besides this, in most compounds with large magnetoresistance, the amplitude of SdH oscillations is much weaker than the background, making it difficult to detect and extract the oscillation signal for analysis. To probe the Fermi surface of SrSn$_4$ we have performed de Haas-van Alphen (dHvA) oscillations measurements. For the first time in SrSn$_4$, **Figure 2**a, **Figure 3**a, and **Figure 3**d display large amplitude dHvA oscillations (*ΔM*) comparable to the background magnetic moment for all three perpendicular crystallographic axes. As **Figure 2**a shows, the oscillations in the B∥b-axis configuration have three frequency (*F*) components: a high-frequency component, a low-frequency component, and a medium-frequency component with a relatively weaker oscillation amplitude.

To separate the three components a deconvolution is performed. In the first step, the highest frequency component is separated from *ΔM* (see **Figure 2**b). The fast Fourier transformation (FFT) spectra in the inset reveal a peak at 234 T. The remaining part of *ΔM* contains the two lower frequency components. The amplitude of the medium-frequency oscillations is weaker at lower temperatures (up to 8 K) but becomes prominent at higher temperatures (above 12 K). This is because the lowest frequency oscillations are suppressed faster with increasing temperatures. A comparable oscillation amplitude of the two lower frequency components at 9 K, 10 K, and 11 K is evident from the residual $\Delta M$ vs $1/B$ plot and FFT spectra in Figure S2, Supporting Information. With further background processing, the stronger lowest frequency component at lower temperatures is separated and plotted in $1/B$ in **Figure 2c**. The FFT spectra in the inset show a frequency peak at 15.5 T. At higher temperatures, above 12 K, the medium frequency component in $\Delta M$ is separated and plotted in **Figure 2e**. The dominance of this medium-frequency component in the oscillating part of $M$ over the 15.5 T frequency component at higher temperatures is reflected in the top-left inset of the figure. The lower-right inset shows a frequency peak at 49 T. The three frequencies correspond to the three extreme cross-sectional areas ($A_F$) of the Fermi pockets normal to the $b$-axis. The values of $A_F$ calculated using the Onsager relationship, $F = \frac{\phi_0}{2\pi^2} A_F$, where $\phi_0$ is the magnetic flux quantum, are listed in **Table 1**. The value of the Fermi wave vector, $k_F$ is obtained by circular approximation of $A_F$.

**Table 1.** B∥b-axis configuration: Frequencies (F), Fermi wave vector ($K_F$), Cross-sectional area ($A_F$), Effective mass of charge carrier ($m^*$), Fermi velocity ($V_F$), Dingle Temperature ($T_D$), Carrier scattering time ($\tau_q$), and Quantum mobility ($\mu_q$). The accuracy of our analysis is 1-2 % for the determination of $F(T)$, $K_F$, and $A_F$, and 5-10 % for the other parameters.

| F (T) | $K_F$ (Å$^{-1}$) | $A_F$ (Å$^{-2}$) | $m^*$ | $V_F$ (10$^5$ m/s) | $T_D$ (K) | $\tau_q$ (s) | $\mu_q (cm^2 V^{-1} s^{-1})$ |
|---|---|---|---|---|---|---|---|
| 234 | 8.4×10$^{-2}$ | 2.2×10$^{-2}$ | 0.260 $m_e$ | 3.7 | 3.2 | 3.75×10$^{-13}$ | 2.53×10$^3$ |
| 15.5 | 2.2×10$^{-2}$ | 1.5×10$^{-3}$ | 0.122 $m_e$ | 2 | 8.3 | 1.46×10$^{-13}$ | 2.30×10$^3$ |
| 49 | 3.8×10$^{-2}$ | 4.6×10$^{-3}$ | 0.083 $m_e$ | 5.4 | 10.1 | 1.20×10$^{-13}$ | 2.64×10$^3$ |

Quantum oscillations are described using the Lifshitz-Kosevich (LK) formula which, in the case of oscillations in magnetization, has the following expression.[51,61]

$$\Delta M \propto -B^{\frac{1}{2}} R_T R_D R_S \sin[2\pi(\frac{F}{B} - \gamma - \delta)] \quad (1)$$

The thermal damping of the oscillation amplitude is described by $R_T = \frac{2\pi^2 k_B T/\hbar\omega_c}{\sinh(2\pi^2 k_B T/\hbar\omega_c)}$, the Dingle temperature ($T_D$) term $R_D = \exp(\frac{2\pi^2 k_B T_D}{\hbar\omega_c})$, and the temperature and magnetic field independent Zeeman spin splitting term $R_S = \cos\left(\frac{\pi g\mu}{2}\right)$. Here, $\omega_c = eB/m^*$ and $\mu = \frac{m^*}{m_0}$, where $m^*$ and $m_0$ are the effective mass of the charge carrier and the free electron mass, respectively. The sine term represents the phase term, where $\gamma = \frac{1}{2} - \frac{\phi_B}{2\pi}$ and $\phi_B$ is the Berry phase. The phase shift $\delta$, which is determined by the dimensionality of FS, is 0 and ± 1/8, respectively for 2D and 3D.

Theoretical fit to thermal damping of oscillation amplitude for 234 T, 49 T, and 15.5 T frequencies, as shown in **Figure 2d** and **Figure** 2e, with $R_T$ provides the effective mass of charge carriers. The values of effective mass and Fermi velocity, $v_F = \frac{\hbar k_F}{m^*}$, are mentioned in **Table 1**. The values of $T_D$ are obtained from the slope of $\ln\left[\frac{\Delta M}{B^{\frac{1}{2}} R_T}\right]$ vs $\frac{1}{B}$ plots which are shown in Figure S3a,b,c, Supporting Information. The quantum relaxation time ($\tau_q$) and quantum mobility ($\mu_q$) of charge carriers are calculated using the relation $\tau_q = \frac{\hbar}{2\pi k_B T_D}$ and $\mu_q = \frac{e\hbar}{2\pi k_B m^* T_D}$, respectively. The extracted values of parameters listed in **Table 1** reveal a significant variation across the Fermi pockets corresponding to the three distinct frequencies. **Table 1** shows the high quantum mobility of charge carriers in SrSn4, which is similar to that observed in other topological electronic materials.[11,51,53,62,63] It is worth noting that the quantum mobility of charge carriers is typically much lower than the classical mobility because of its sensitivity to small-angle scattering. Following the convention which assigns the valleys in $\Delta M$ to $n - \frac{1}{4}$ and peaks to $n + \frac{1}{4}$, where the Landau level index (LL index), $n$, is an integer, Berry's phase is calculated from the intercept of the *LL index* vs $\frac{1}{B}$ plot, as is shown in the inset of **Figure 2**f for the 49 T.[51,63,64] The extracted intercept of 0.23 corresponds to Berry phase: $\phi_B = 0.71\,\pi$ for $\delta = +\frac{1}{8}$ and $\phi_B = 0.21\,\pi$ for $\delta = -\frac{1}{8}$. Berry phases closer to zero are observed for the other two frequencies.

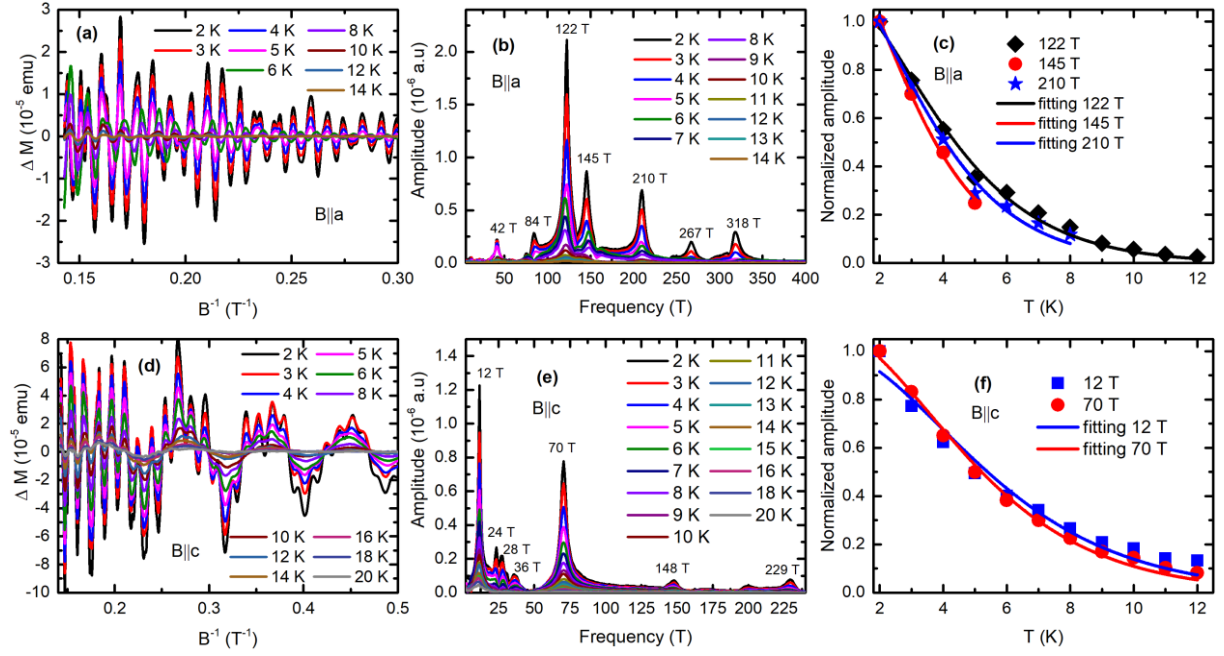

**Figure 3**: dHvA oscillations for magnetic field applied along *a*- and *c*-axis directions in the SrSn4 single crystals. a) $\Delta M$ after background subtraction plotted in $\frac{1}{B}$ for magnetic field parallel to *a*-axis configuration. b) FFT spectra of $\Delta M$ vs $\frac{1}{B}$. c) Thermal damping of oscillations amplitudes and theoretical fitting for the three prominent peaks, 122 T, 145 T, and 210 T in the FFT spectra. The same for the other frequency peaks are not shown. d) $\Delta M$ vs $\frac{1}{B}$ for $B$ parallel to the *c*-axis configuration. e) The corresponding FFT spectra. f) Theoretical fit to the thermal damping of oscillations amplitudes for the two most prominent peaks. The same for the other frequency peaks are not shown.

**Table 2.** B∥a-axis configuration: Frequencies (F), Fermi wave vector ($K_F$), Cross-sectional area ($A_F$), Effective mass of charge carrier ($m^*$), and Fermi velocity ($V_F$). The accuracy of our analysis is 1-2 % for the determination of $F\,(T)$, $K_F$, and $A_F$, and 5-10 % for the other parameters.

| $F\,(T)$ | $K_F\,(\text{Å}^{-1})$ | $A_F\,(\text{Å}^{-2})$ | $m^*$ | $V_F\,(10^5\ \text{m/s})$ |
|---|---|---|---|---|
| 42 | $3.5 \times 10^{-2}$ | $3.9 \times 10^{-3}$ | $0.038\,m_e$ | 10.7 |
| 84 | $5.0 \times 10^{-2}$ | $7.8 \times 10^{-3}$ | $0.071\,m_e$ | 8.2 |
| 122 | $6.0 \times 10^{-2}$ | $1.1 \times 10^{-2}$ | $0.068\,m_e$ | 10.3 |
| 145 | $6.6 \times 10^{-2}$ | $1.4 \times 10^{-2}$ | $0.086\,m_e$ | 8.9 |
| 210 | 7.9×10⁻² | 2.0×10⁻² | 0.076 $m_e$ | 12.1 |
| 267 | 9.0×10⁻² | 2.5×10⁻² | 0.106 $m_e$ | 9.8 |
| 318 | 9.8×10⁻² | 3.0×10⁻² | 0.107 $m_e$ | 10.6 |

The dHvA oscillations in magnetization for the other two crystallographic configurations, B∥a-axis and B∥c-axis are shown in **Figure 3**, and the results are summarized in **Table 2** and **Table 3**. FFT of $\Delta M$ for B∥a-axis gives us frequency peaks at 42 T, 84 T, 122 T, 145 T, 210 T, 267 T, and 318 T. Some of the frequencies, for example, 84 T and 210 T, appear to be higher harmonics of the fundamental frequency, 42 T. However, the amplitude of higher harmonic quantum oscillations is usually significantly lower than that of the fundamental frequency. This trend is not observed in the present case, as shown in **Figure 3**b. Furthermore, the effective masses of charge carriers and Fermi velocities are significantly different for Fermi pockets corresponding to these frequencies. Thus, we consider all of them as fundamental frequencies, acknowledging the presence of some degree of error, although not significant. The $\Delta M$ for B∥c-axis configuration and FFT analysis reveal frequency peaks at 12 T, 24 T, 28 T, 36 T, 70 T, 148 T, and 229 T (**Figure 3**d and **Figure 3**e). It is noted that 24 T and 36 T appear to be higher harmonics of the 12 T frequency, as they are much weaker compared to the fundamental frequency, as evident in **Figure 3**e. Additionally, the listed values of $m^*$ and $v_F$ for 12 T, 24 T, and 36 T frequencies in **Table 3** are similar. The values of $m^*$ and $v_F$ for other frequency oscillations are also listed in **Table 3**. It is apparent that $m^*$ of charge carriers in the plane perpendicular to the a- and c- axes is lighter compared to the effective masses in the plane perpendicular to the b-axis. In general, we find that in SrSn$_4$ for all the Fermi pockets $m^*$ is very low and comparable to values observed in well-known topological semimetals such as Cd$_3$As$_2$, NbAs, NbP, TaAs, and ZrSiS.[11,65-67]

**Table 3.** B ∥ c-axis configuration: Frequencies (F), Fermi wave vector ($K_F$), Cross-sectional area ($A_F$), Effective mass of charge carrier ($m^*$), and Fermi velocity ($V_F$). The accuracy of our analysis is 1-2 % for the determination of F (T), $K_F$, and $A_F$, and 5-10 % for the other parameters.

| F (T) | $K_F$ (Å⁻¹) | $A_F$ (Å⁻²) | $m^*$ | $V_F$ (10⁵ m/s) |
|---|---|---|---|---|
| 12 | 1.9×10⁻² | 1.1×10⁻³ | 0.037 $m_e$ | 5.9 |
| 24 | 2.7×10⁻² | 2.2×10⁻³ | 0.052 $m_e$ | 5.9 |
| 28 | 2.9×10⁻² | 2.6×10⁻³ | 0.048 $m_e$ | 7.0 |
| 36 | 3.3×10⁻² | 3.4×10⁻³ | 0.042 $m_e$ | 9.1 |
| 70 | 4.6×10⁻² | 6.6×10⁻³ | 0.042 $m_e$ | 12.8 |
| 148 | 6.7×10⁻² | 1.4×10⁻² | 0.040 $m_e$ | 19.1 |
| 229 | 8.3×10⁻² | 2.2×10⁻² | 0.046 $m_e$ | 21.1 |

## 2.3. Magnetotransport and Shubnikov-de Haas (SdH) oscillations study

Resistivity measurements ($\rho_{xx}$) in Figure 4a in the transverse magnetic field configuration with B∥b and the current I∥a, reveals a large magnetoresistance given by $TMR = \frac{\rho_{xx(B)} - \rho_{xx}(0)}{\rho_{xx}(0)} \times 100\% \approx 1200$ % at 5 K and 14 T. The field dependence is slightly sublinear with no detectable SdH oscillations. However, when we rotate our sample to align the *c*-axis along the magnetic field while keeping the current direction the same, we observe prominent oscillations in the resistivity along with a large $TMR$. **Figure 4**b,c shows that the magnitude of $TMR$ for B∥c is similar to that for B∥b. However, in contrast to B∥b, $TMR$ for B∥c exhibits a linear behavior over the entire field range in addition to the SdH oscillations. The oscillating part in $\rho_{xx}$, $\Delta\rho_{xx}$ is separated from the background resistivity and plotted in **Figure 4**d as a function of $\frac{1}{B}$ at different temperatures up to 10 K. The FFT spectra in **Figure 4**e display a single frequency peak at 70 T. The dHvA oscillations for the same B∥c configuration, shown in Figure 3(e), also feature a pronounced FFT peak exactly at the 70 T frequency. Similar to dHvA oscillations, SdH oscillations in resistivity are described by the Lifshitz-Kosevich (LK) formula: $\Delta\rho_{xx} \propto R_T R_D \cos 2\pi(\frac{F}{B} + \frac{1}{2} - \frac{\phi_B}{2\pi} - \delta)$.[61,68] The theoretical fit to the thermal damping of oscillation amplitude with the expression $R_T$ is shown in the inset of **Figure 4**e and the effective mass $m^*$ of the charge carriers corresponding to the 70 T Fermi pocket is determined. The Dingle temperature, $T_D$, is determined from the slope of $\ln\left[\frac{\Delta\rho_{xx}}{R_T}\right]$ vs $\frac{1}{B}$ plot which is shown in Figure S4, Supporting Information. Quantum relaxation time and quantum mobility are obtained using the previously mentioned expressions and listed in **Table 4**. We observe that $m^*$ of charge carriers corresponding to 70 T frequency obtained in the SdH oscillations study is almost three times heavier than the value obtained from the dHvA oscillations analysis. It is common in the literature to find heavier $m^*$ values from SdH oscillations possibly due to its complex dependence on scattering mechanisms.[56,57,69-71] The most fascinating outcome of the present SdH oscillation study is the detection of a Berry's phase close to $\pi$, as seen from the LL-index plot in **Figure 4**f. Following convention,[68,72] we assign valley positions in $\Delta\rho_{xx}$ as integer index, $n$, and peak positions as half-integer index, $n + \frac{1}{2}$, and plot the LL index against $\frac{1}{B}$. The linear extrapolation of the plot yields an intercept of approximately 0.53, which is very close to the expected value of 0.5 for a topological electronic band exhibiting a nontrivial $\pi$ Berry's phase. This intercept of 0.53 corresponds to $\phi_B = 1.06\,\pi$, which falls well within the range of $\pi \pm \frac{1}{8} \times 2\pi$ for a 3D topological electronic system.

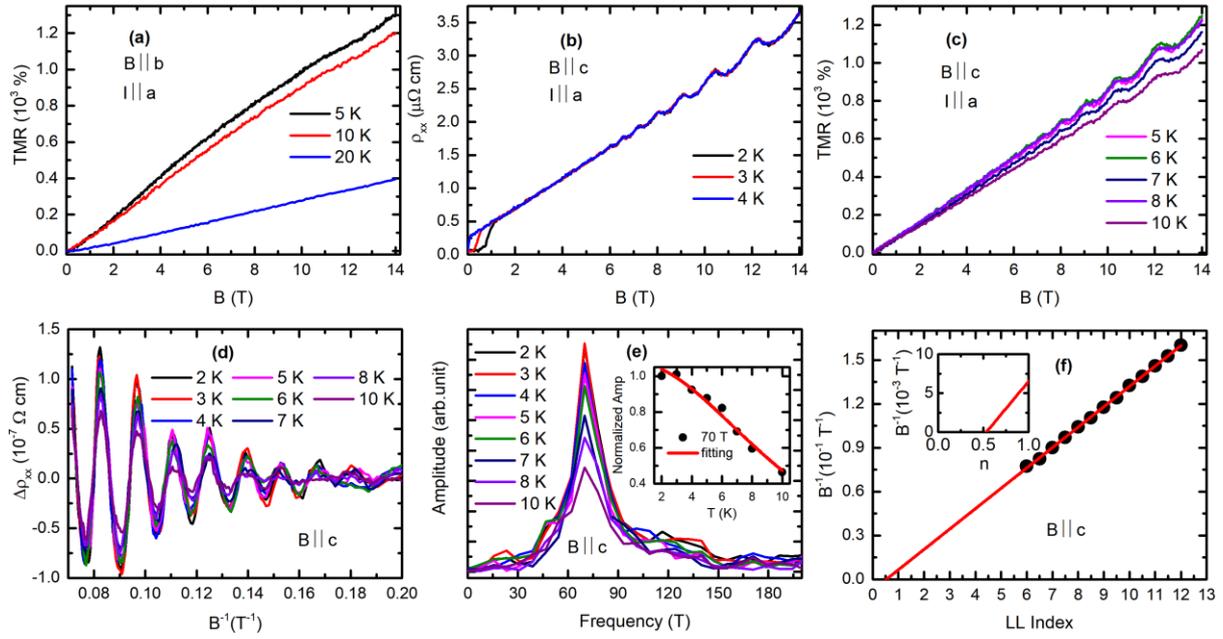

**Figure 4:** Magnetotransport and SdH oscillations in the SrSn4 single crystals. a) Transverse magnetoresistance (*TMR*), defined as $\frac{\rho_{xx(B)} - \rho_{xx}(0)}{\rho_{xx}(0)} \times 100\%$, with increasing magnetic field when *B*||*b*-axis and current, *I*||*a*-axis. b) ρxx (B) at representative temperatures below $T_C$ when the magnetic field is along the *c*-axis and the current is along the *a*-axis. c) *TMR* for the same experimental configuration above $T_C$. d) Oscillating part of resistivity after background subtraction (Δρxx) as a function of $\frac{1}{B}$. e) FFT of Δρxx vs $\frac{1}{B}$. Inset shows the temperature dependence of oscillations amplitude and theoretical fit with the LK formula. f) The Landau level index plot of the oscillations involves considering valley positions as integer indexes and peak positions as half-integer indexes, following convention.

**Table 4.** Parameters calculated from SdH oscillations analysis for B ∥ c-axis configuration: Frequencies (F), Fermi wave vector ($K_F$), Cross-sectional area ($A_F$), Effective mass of charge carrier ($m^*$), and Fermi velocity ($V_F$), Dingle Temperature ($T_D$), Carrier scattering time ($\tau_q$), and Quantum mobility ($\mu_q$). The accuracy of our analysis is 1-2 % for the determination of *F (T)*, $K_F$, and $A_F$, and 5-10 % for the other parameters.

| F (T) | $K_F$ (Å$^{-1}$) | $A_F$ (Å$^{-2}$) | $m^*$ | $V_F$ ($10^5$ m/s) | $T_D$ (K) | $\tau_q$ (s) | $\mu_q (cm^2 V^{-1} s^{-1})$ |
|---|---|---|---|---|---|---|---|
| 70 | 4.6×10$^{-2}$ | 6.6×10$^{-3}$ | 0.12 $m_e$ | 4.4 | 14.8 at 5 K | 8.2×10$^{-14}$ | 1.2×10$^3$ |

To investigate the anisotropic magnetotransport, we conducted $\rho_{xx}$ measurements keeping *I* parallel to the *a*-axis and varying angle *θ* between *B* and the *b*-axis, as shown in **Figure 5**a. Angle dependence of Resistivity $\rho_{xx}(\theta)$ at 2 K and 10 K is shown in **Figures 5**b,c at representative field strengths of 5 T, 10 T, and 14 T. In **Figure 5**c, we normalized $\rho_{xx}(\theta)$ at 10 K into *TMR* (%) for better representation of the electronic property of technological significance. $\rho_{xx}(\theta)$ at 2 K and *TMR* (*θ*) at 10 K show four-fold symmetry with the maximum value occurring close to $\theta = 45^0$ and the minimum value close to $\theta = 90^0$, taking into account a few degrees of error in our sample rotator setup. Even at 10 K, *TMR* is as high as 1850% at 14 T. Given the approximately 10% higher *TMR* observed at 5 K, as shown in **Figure 4**a and **Figure 4**c, it is reasonable to infer that the *TMR* in the current compound exceeds 2000% at 14 T. The ratio of $\frac{\rho_{xx}(maximum)}{\rho_{xx}(minimum)}$ at 2 K is approximately 2 and the ratio of $\frac{TMR(maximum)}{TMR(minimum)}$ at 10 K is approximately 1.8 at 14 T. As the magnetic field decreases, these ratios decrease and reach a value of approximately 1.5 at 5 T. A closer look at the four-fold symmetric angular variation of $\rho_{xx}(\theta)$ and *TMR* (*θ*) reveals a superimposed modulation that creates shallow peaks and valleys, particularly at higher fields. We infer that this is due to the large SdH oscillation amplitude at high fields and its angular dependence as illustrated in **Figure 4**a and **Figure 4**b.

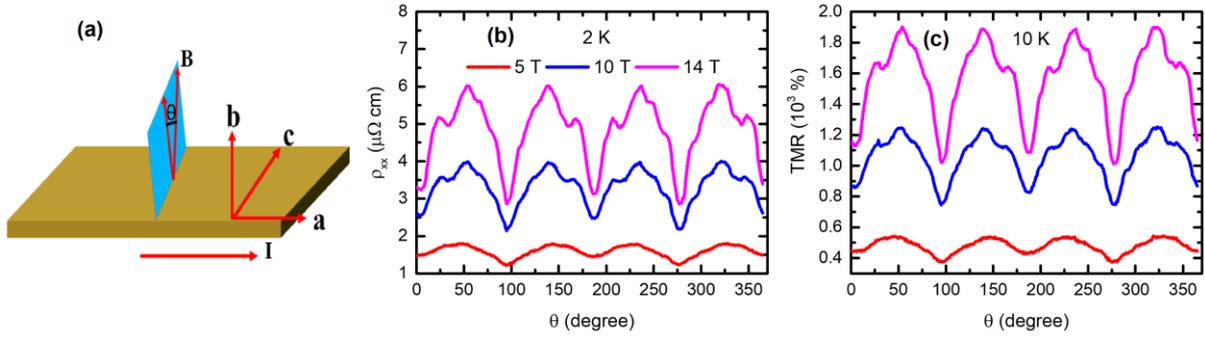

**Figure 5**: Crystallographic direction-dependent magnetotransport in the SrSn$_4$ single crystals. a) Schematic of experimental configuration for the crystallographic direction-dependent magnetotransport experiments. b) $\rho_{xx}$ as a function of the angle ($\theta$) between magnetic field direction and *b*-axis at the temperature, 2 K at some representative field strengths above $H_{c2}$. c) Angle dependence of $TMR$ at 10 K, a representative temperature well above $T_C$ of SrSn$_4$.

## 2.4. Theory calculations

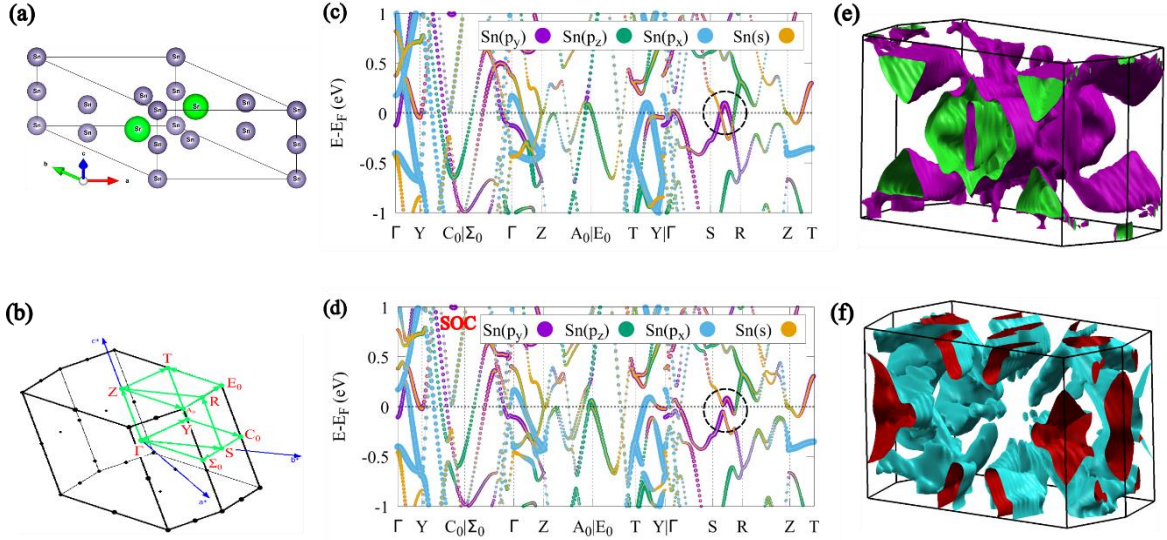

**Figure 6:** Band structure theory calculations of SrSn$_4$. a) A primitive cell of SrSn$_4$ belonging to the space group Cmcm (No. 63) is shown here. b) The corresponding Brillouin zone is plotted highlighting the high-symmetry points and directions where the band structures are shown. c and d) Calculated band structures are shown along with the orbital weight density (represented by filled circles' size) of Sn s and p orbitals. Panel c) and d) correspond to the same calculation without and with SOC. The dashed circle along the S-R direction emphasizes the gap opening due to SOC. e) and f) The two large Fermi surfaces of SrSn$_4$. The other two small Fermi surfaces are shown in Figure S6, Supporting Information.

The crystallographic structure of SrSn$_4$ is classified under the orthorhombic crystal system with the space group $Cmcm$ (number 63) and point group D$_2$h (number 8), see **Figure 6**a. The lattice parameters of the relaxed primitive cell are $a = b = 9.1095$ Å and $c = 7.1924$ Å, while those for the formula unit (doubling the number of atoms) are $a = 4.6179$ Å, $b = 17.3720$ Å, and $c = 7.0602$ Å. Figure 6b displays the Brillouin zone, highlighting the high-symmetry points and directions where the band structures are shown. The band structure is compared without and with spin-orbit coupling (SOC), in **Figure 6**c and **Figure 6**d, exhibiting an inverted band gap opening mainly along the S – R direction. This finding is consistent with previous band structure predictions, which identify SrSn$_4$ as a 3D topological insulator (TI) when taking into account SOC, employing the theory of topological quantum chemistry.[1-3] Although SrSn$_4$ has been classified as a topological insulator (TI), its density of states at the Fermi level is not zero due to the presence of metallic trivial bands. SrSn$_4$ continues to exhibit metallic behavior with SOC with considerable dispersion and three-dimensionality. We also compare the contributing orbital weight for the low-energy bands, exhibiting dominant contributions from the Sn s and p - orbitals, while the Sr atomic weights largely departed from the Fermi level region. The band inversion in the high-symmetry S-R direction, highlighted by the dashed circle in **Figure 6**c, occurs due to the anisotropic hybridization between the Sn $p_z$ and $s$ - orbitals. This mixing of orbitals with opposite parity introduces chirality in the band structure along the z-direction, causing a novel mechanism of orbital selective topology[73] beyond the quintessential helicity-driven topology due to SOC. The topology of the band structure in the normal state is expected to be retained in the Bogolyubov quasiparticle spectrum in the superconducting state due to the theory of adiabatic continuity.[73]

## 2.5. Discussion

In our dHvA oscillations study, we have identified three Fermi surface cross-sectional areas perpendicular to the b-axis, and eight such cross-sections in the plane perpendicular to the a-axis. For the configuration, $B\|c$, we have discerned five cross-sectional areas corresponding to five fundamental frequencies. The calculated Fermi surfaces, as shown in **Figure 6** and Figure S6, Supporting Information, consist of two small, disconnected, ellipsoidal Fermi pockets, while the other two Fermi surfaces are large and strongly anisotropic. This strong three-dimensionality of the Fermi surfaces results in multiple quantum oscillation frequencies, depending on the field direction, precluding us from theoretically estimating their areas perpendicular to the direction of the sweeping magnetic field. Moreover, real materials are inevitably subject to uncontrolled doping during synthesis; our sample, as discussed in the Supporting Information and illustrated in Figure S5, is hole-doped, potentially leading to contributions from other bands not accounted for in theoretical calculations at the Fermi level. A thorough angular progression mapping of frequency peaks by varying the angle between the crystallographic axes and the magnetic field direction at small intervals is needed to obtain the three-dimensional geometry of individual Fermi pockets. So, a more comprehensive experimental exploration of the Fermi surface geometry of SrSn$_4$ requires further research.

Nevertheless, the most promising outcome of our quantum oscillation experiments is the detection of a topological electronic band, supported by band structure theory calculations. The remarkably low effective mass of charge carriers observed in SrSn$_4$, along with a high quantum mobility, hints at the existence of a topologically nontrivial electronic state. The SdH oscillations study unveils a nontrivial π-Berry phase, a widely recognized hallmark of a topologically nontrivial electronic state in a material. We assert that at least one electronic band corresponding to the 70 T frequency observed for B||c is topological in nature. While thousands of topological electronic materials have been predicted to exist, and hundreds have been experimentally confirmed, only a few exhibit both topological electronic bands and superconductivity, making them promising candidates for topological superconductivity.[74-78] Despite superconductivity being present in those few topological compounds, the superconducting transition temperature, $T_c$, is typically quite low, often well below the boiling point of helium. In this respect the experimental identification of a material with topological states and $T_c > 4.2$ K at ambient pressure is extremely rare.[24,25] The present study identifies SrSn$_4$ as one such rare material. When comparing SrSn$_4$ to other known superconducting topological compounds with significant magnetoresistance in **Table 5**, the exceptionality of SrSn$_4$ as the only topological compound with large magnetoresistance and $T_C$ above the boiling point of helium at ambient pressure becomes apparent. For other topological compounds with $T_C$ higher than that of SrSn$_4$ we find that the magnetoresistance is either negligible or not reported. For example, the magnetoresistance in NaAlSi is only 7% at low temperature and a magnetic field of 4 T.[26] Furthermore, the highly linear magnetoresistance observed for the B∥$c$ and I∥$a$ configuration, along with the crystallographic direction-dependent anisotropic magnetotransport (with a $\frac{TMR(maximum)}{TMR(minimum)}$ ratio of approximately 2), highlights its technological importance.

**Table 5.** Experimentally verified candidate topological electronic materials with large magnetoresistance and superconductivity.

| Material | Superconducting $T_c$ | Magnetoresistance | Reference |
| --- | --- | --- | --- |
| MoTe$_2$ | 0.1 K | 4000 % at 2 K, 8.5 T | [34, 35] |
| γ-PtBi$_2$ | 0.5 K | 6000 % at 1 K, 20 T | [37,38] |
| α-Ga | 0.9 K | 1.66×10$^6$ % at 2 K, 9 T | [36] |
| Au$_2$Pb | 1 K | 100 % at 1 K, 15 T | [32] |
| YPtBi | 1 K | 200 % at 4 K, 14 T | [31] |
| CaSb$_2$ | 1.7 K | 2500% at 3 K and 9 T | [42] |
| PbTe$_2$ | 2 K | 900 % at 1.5 K, 30 T | [33] |
| AuSn$_4$ | 2.4 K | 550% at 2 K and 9 T | [39, 40] |
| PtPb$_4$ | 2.7 K | 350 % at 5 K and 9 T | [41] |
| SnTaS$_2$ | 2.8 K | ~ 250% at 1.8 K and 10 T | [43] |
| PbTaSe$_2$ | 3.8 K | 150 % at 5 K, 9 T | [28,29,30] |
| SrSn$_4$ | 4.8 K | 1850 % at 10 K, 14 T | Present compound |

Although there have been reports on the deviation from conventional BCS superconductivity in SrSn$_4$,[23] the material lacks comprehensive research so far. This could be due to its oxidizing nature in ambient air, complex Sn-flux growth which leads to thin Sn patch residue on the surface of the as grown crystals, and multiband contributions in electronic transport as evident from previous theoretical analysis.[23] We believe that our discovery of topological electronic states and exciting magnetotransport properties, facilitated through an optimized clean crystal growth recipe, will stimulate the community to conduct dedicated investigations aimed at understanding the superconductivity in SrSn$_4$. In future research, considering both the idea of multiband superconductivity, as proposed by X. Lin et al.,[23] and the presence of topological states could be beneficial in establishing the nature of superconductivity in SrSn$_4$.

## 3. Conclusions

To conclude, we have synthesized high-quality single crystals of SrSn$_4$ using an optimized growth technique. The crystals are Sn-flux-free and exhibit large-amplitude quantum oscillations in both magnetization and resistivity measurements. Our quantum oscillations study unveils multiple Fermi pockets characterized by very low effective mass and high quantum mobility of the charge carriers. The detection of a $π$-Berry phase confirms the presence of a nontrivial topological state. Magnetotransport experiments identify SrSn$_4$ as the only topological compound with a superconducting $T_c$ above the boiling point of helium at ambient pressure, which shows a remarkably high magnetoresistance. The observation of a highly linear field-dependent magnetoresistance underscores its technological importance. The high superconducting $T_c$ of SrSn$_4$ in particular, exceeding the boiling point of helium, holds promise to broaden the research horizon across different experimental techniques, facilitated by readily available 4 K cryogenic setups.

## 4. Experimental and Computational details

*Sample preparation.* Single crystals of SrSn$_4$ were grown by Sn flux method, with reference to the Sr–Sn binary phase diagram reported in literature.[79] Elemental Sr (distilled dendritic pieces, 99.8 %, Alfa Aesar) and Sn (shots, 5N, Alfa Aesar) at a molar ratio of 2 : 98, were loaded into frit-disc alumina crucibles, and sealed in a quartz ampule under vacuum (~1 × 10$^{-5}$ mbar). A thin layer of quartz wool was placed between the frit-disc and the bottom alumina crucible, without touching the starting elements, to provide cushioning for the as-grown crystals during centrifugation. All steps of materials handling were performed in an Ar glove box with O$_2$ and H$_2$O concentrations of ~ 0.1 ppm. The sealed ampule was placed vertically in a box furnace and heated to 800 °C at a rate of 20 °C h$^{-1}$. The ampule was kept at 800 °C for 48 hours, then cooled down to 370 °C at a rate of 10 °C h$^{-1}$, followed by very slow cooling at 1 °C h$^{-1}$ to 260 °C. The ampule was kept at that final temperature for 72 hours. It was then flipped upside down and within seconds transferred to a centrifuge to remove the excess Sn flux at 3000 RPM. The ampule was again placed in the box furnace and slowly heated to 280 °C, kept there for 12 hours, and then centrifuged a second time to ensure the removal of excess flux from the crystals surfaces. Finally, the as grown crystals were placed on a hotplate inside the Ar glove box at 240 °C, i.e., just above the melting point of tin. Then a fine cotton bud was used to wipe the surfaces of the crystals to ensure the removal of any thin layer of tin that may still exist after the second centrifuge. Crystals were stored in the Ar glove box. Before any measurement or characterization, a gentle polishing of the as-grown crystals using 800 mesh size sanding paper was performed to guarantee the complete absence of residual Sn layers as well as surface oxidation. The described growth recipe is the optimal method for obtaining extremely high quality samples. The cooling rate from 370 °C to 260 °C was varied between 3 °C h$^{-1}$ and 1 °C h$^{-1}$, the molar mixing ratio of Sr and Sn was varied from 3.5 : 96.5 to 2 : 98, the heating temperature for the second centrifuge was tuned between 300 to 280 °C, and the quartz ampule was sealed both in vacuum (~1 × 10$^{-5}$ mbar) and ~ 0.3 bar Ar (5N) in several other batches of sample growth. Based on various criteria, including regular morphology, absence of Sn flux residue on the crystal surfaces, higher superconducting transition temperature, presence of high-quality quantum oscillations, and a large residual resistivity ratio (RRR = $\frac{\rho(300\,K)}{\rho(5\,K)}$), we identified the highest quality crystals.

*Measurement details.* The single crystal X-ray diffraction characterization was performed at 100 K on a Rigaku Synergy S 4-circle Kappa System with a microfocus sealed-tube Ag-source and a Dectris Pilatus 3R 300K CdTe detector. CrysAlisPro package was used for data collection and processing. 'SHELXL was used for structure refinement. The output crystallographic information file (cif) is provided in additional information files.

The hard X-ray photoelectron spectroscopy (HAXPES) experiment was carried out at beamline P22 at PETRA III, DESY, Hamburg, Germany. A post-monochromator was used to improve the resolution and stability of the photon beam. The overall energy resolution, including the source and the analyzer contribution, was 0.3 eV, as measured from the Au Fermi edge that was in electrical contact with the specimen. Details related to the beamline and the end station can be found in Schlueter et al.[80] The SrSn$_4$ single crystal was scraped in-situ at a pressure of 1×10$^{-8}$ mbar using a diamond file and was immediately transferred to the analysis chamber having a pressure of 2×10$^{-9}$ mbar. The non-linear least square curve fitting has been performed using a DS line shape and Tougaard background, as in our earlier work.[81]

The magnetization measurements were performed in a 7-T superconducting quantum interference device (SQUID) vibrating-sample magnetometer (Quantum Design). A standard quartz sample holder was used for mounting the sample with GE varnish. The dc electrical transport measurements were performed using the standard four-probe technique in a 14 T physical property measurement system (PPMS) from Quantum Design. Electrical contacts were made using conductive silver paste and gold wire. A horizontal sample rotator was employed to perform the crystallographic direction-dependent resistivity measurements.

*Band Structure Calculations*. We performed density-functional theory (DFT) simulations using the Vienna Ab initio Simulation Package (VASP).[82,83] For the exchange correlation between electrons, we employed the generalized gradient approximation (GGA) parametrized by Perdew-Burke-Ernzerhof (PBE).[84] The projector-augmented wave (PAW) approach was used to characterize the interaction between the core and valence electrons.[85,86] We set the plane-wave energy cut-off to 500 eV and sampled the Brillouin zone (BZ) using the Monkhorst-Pack technique with a 11×11×7 k-point mesh,[87] for the structural relaxation of the primitive cell. The structural relaxation was carried out until the forces acting on each atom were less than 0.0001 eV/Å. The convergence threshold for energy in the electronic self-consistent cycle was set to $10^{-8}$ eV. We also performed the spin-orbital coupling (SOC) calculations.

## Acknowledgements

A. K. P. acknowledges the postdoctoral fellowship support from the Council for Higher Education, Israel, through the 'Study in Israel' program, and infrastructure and experimental facilities support from the Weizmann Institute of Science, Rehovot, Israel. A. K. P. acknowledges support from the Department of Science and Technology, India, through the Inspire Faculty Fellowship program (project code SP/DSTO-21-0219), and the support from the current host institution, the Indian Institute of Science, Bangalore, India. M. H. acknowledges funding support from the Helmsley Charitable Trust grant #2112-04911. A. K. P. and M. H. acknowledge various supports from Prof. Eli Zeldov and Condensed Matter Physics Department, Weizmann Institute of Science, Rehovot, Israel. A. K. P. acknowledge the assistance of Dr. Gregory Leitus and Dr. Linda J. W. Shimon from the X-ray Crystallography Laboratory, Chemical Research Support Unit, Weizmann Institute of Science, in conducting the single-crystal X-ray diffraction experiment and analysis. R.S. acknowledges the Science and Engineering Research Board (SERB), Government of India, for providing the NPDF fellowship with grant number PDF/2021/000546. T.D. acknowledges funding from Core Research Grant (CRG) of S.E.R.B. (CRG/2022/003412-G) and benefited from the computational funding from the National Supercomputing Mission (DST/NSM/RDHPC-Applications/2021/39), both are under the Department of Science and Technology, India. MB and SRB are thankful to A. Gloskovskii for help during the HAXPES experiment. MB and SRB gratefully acknowledge the financial support from the Department of Science and Technology, Government of India, within the framework of the India@DESY collaboration. The SERB project CRG/2023/001719 provided MB with a research fellowship.

## Conflict of Interest

The authors declare no conflict of interest.

## Data Availability Statement

The data that support the findings of this study are available from the corresponding author upon reasonable request.

## Authorship contribution statement

A. K. P.: Conceptualization, Materials Synthesis, Magnetic and transport measurements, Formal analysis and data interpretation, Research collaboration, Writing - original draft

R. S.: Band structure theory calculations and writing

M. B.: HAXPES measurements and analysis, and writing

M. H.: Resources, Experimental facility design

T.D.: Band structure theory calculations and writing

S. R. B.: HAXPES measurements and analysis, and writing

All authors contributed to review and editing

**Supporting Information for "Experimental detection of topological electronic state and large linear magnetoresistance in SrSn$_4$ superconductor"**

Arnab Kumar Pariari[1,2,*], Rajesh O Sharma[1], Mohammad Balal[3], Markus Hücker[2], Tanmoy Das[1], Sudipta Roy Barman[3]

[1] Department of Physics, Indian Institute of Science, CV Raman Rd, Bengaluru, Karnataka 560012

[2] Department of Condensed Matter Physics, Weizmann Institute of Science, 234 Herzl Street, POB 26, Rehovot 7610001 Israel

[3] UGC-DAE Consortium for Scientific Research, Khandwa Road, Indore 452001, Madhya Pradesh, India

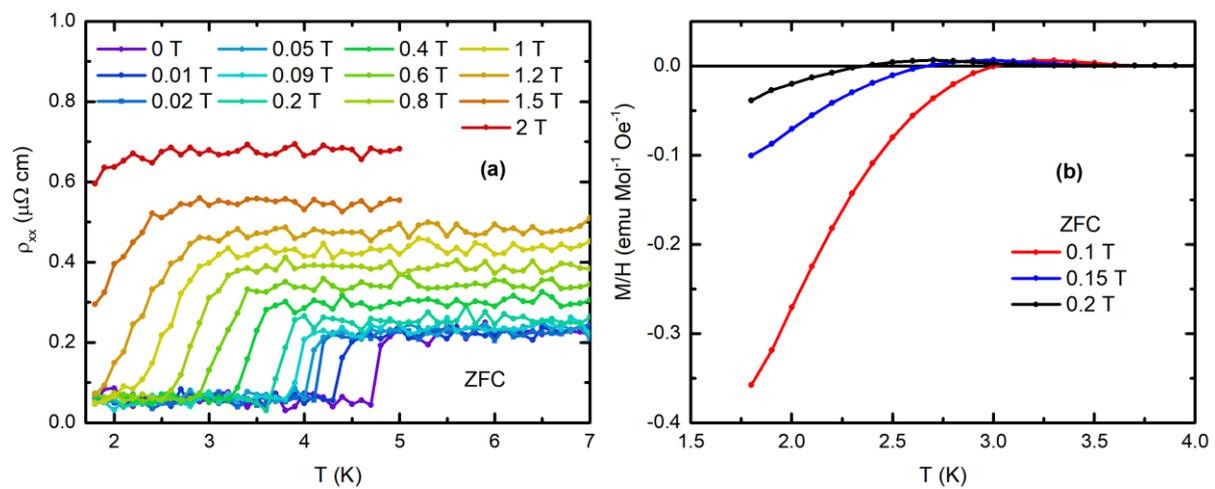

**Supplementary Figure 1**: (a) $\rho_{xx}$ (T) in a narrow temperature range across the superconducting transition at various external magnetic field strengths. (b) Zero field cooled dc magnetic susceptibility (M/H) at some higher representative magnetic fields for B∥b configuration.

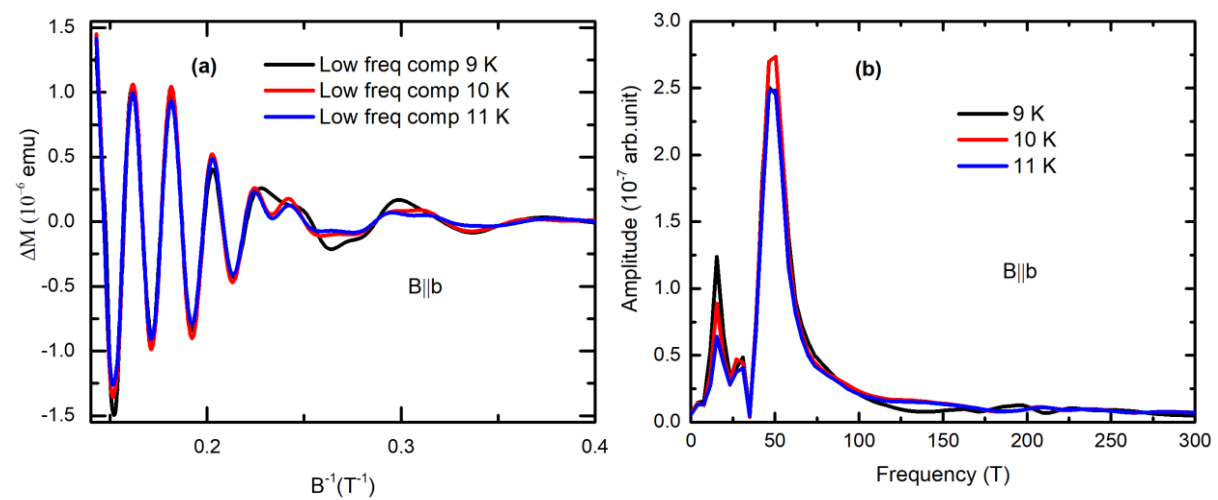

**Supplementary Figure 2**: (a) $\Delta M$ after background subtraction and separation of the high-frequency component, 234 T, for B∥b at 9, 10, and 11 K. (b) The corresponding FFT spectra at these temperatures.

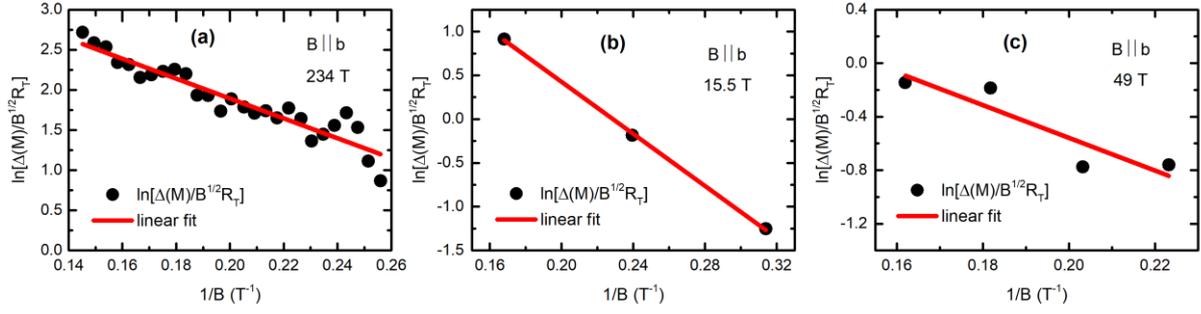

**Supplementary Figure 3**: (a), (b), (c) Dingle temperature plots of the three FFT frequencies 234 T, 15.5 T, and 49 T, respectively in the dHvA oscillation study for the B∥b-axis configuration.

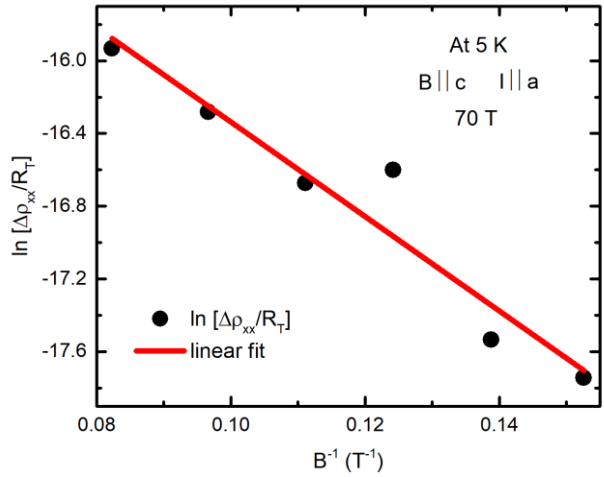

**Supplementary Figure 4**: Dingle temperature plot for the SdH oscillations in transport.

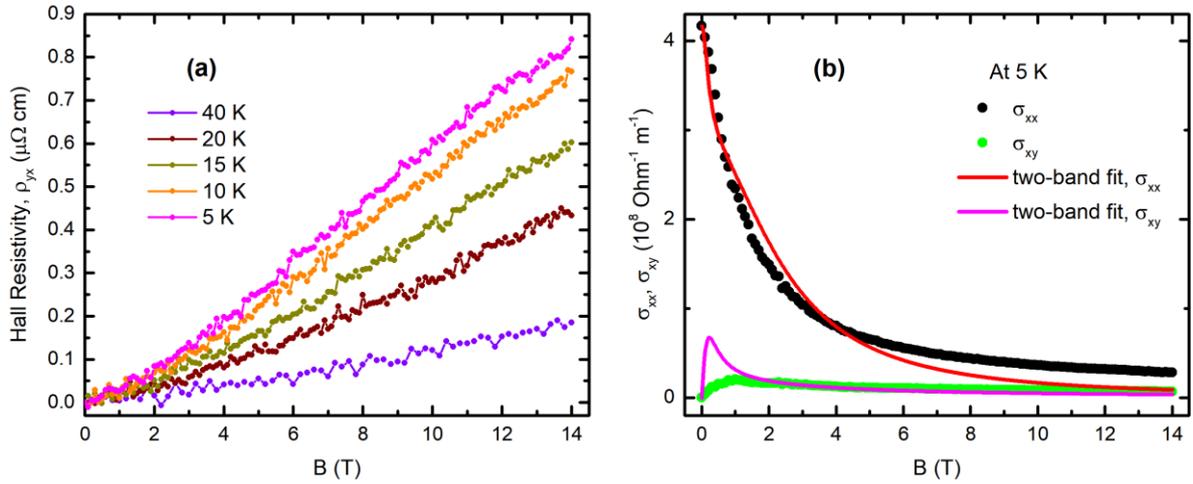

**Supplementary Figure 5**: (a) Hall resistivity, $\rho_{yx}$ vs magnetic field, $B$. (b) Two-band global fitting of $\sigma_{xx}(B)$ and $\sigma_{xy}(B)$ at 5 K with the expressions, $\sigma_{xx}(B) = [\frac{n_h \mu_h e}{1+(\mu_h B)^2} + \frac{(\sigma_{xx}(0) - n_h \mu_h e)}{1+(\mu_e B)^2}]$ and $\sigma_{xy}(B) = [\frac{n_h \mu_h^2}{1+(\mu_h B)^2} - \frac{n_e \mu_e^2}{1+(\mu_e B)^2}]eB$. $\sigma_{xx}(B)$ and $\sigma_{xy}(B)$ are calculated using tensor conversion as follows: $\sigma_{xx}(B) = \frac{\rho_{xx}}{\rho_{xx}^2 + \rho_{yx}^2}$ and $\sigma_{xy}(B) = \frac{\rho_{yx}}{\rho_{xx}^2 + \rho_{yx}^2}$.

The positive Hall resistivity, $\rho_{yx}(B)$, as shown in Supplementary Figure 5(a) suggests that the hole is the majority charge carrier in the present compound. The nonlinear $B$ dependence of $\rho_{yx}$ indicates contributions from multiple bands in transport. The experimental electrical conductivity, $\sigma_{xx}(B)$ and Hall conductivity, $\sigma_{xy}(B)$ are fitted simultaneously with the following two-band model expressions,

$\sigma_{xx}(B) = [\frac{n_h \mu_h e}{1+(\mu_h B)^2} + \frac{(\sigma_{xx}(0) - n_h \mu_h e)}{1+(\mu_e B)^2}]$ and $\sigma_{xy}(B) = [\frac{n_h \mu_h^2}{1+(\mu_h B)^2} - \frac{n_e \mu_e^2}{1+(\mu_e B)^2}]eB$. Here, $n_h$ and $\mu_h$ are density and mobility of hole-type charge carriers, and $n_e$ and $\mu_e$ are density and mobility of electron-type charge carrier, respectively. Supplementary Figure 5(b) displays the red line and magenta line as the best fit achievable from the two-band model, revealing a significant deviation from the experimental data. Our investigation of dHvA oscillations, along with theoretical calculations, reveals that the Fermi surface of SrSn$_4$ consists of multiple Fermi pockets, each with unique transport parameters, shapes, and sizes. So, the deviation of the theoretical fit from experiment data suggests that the two-band model is not adequate to reproduce the transport results and accurately determine the carrier density and mobility of charge carriers.

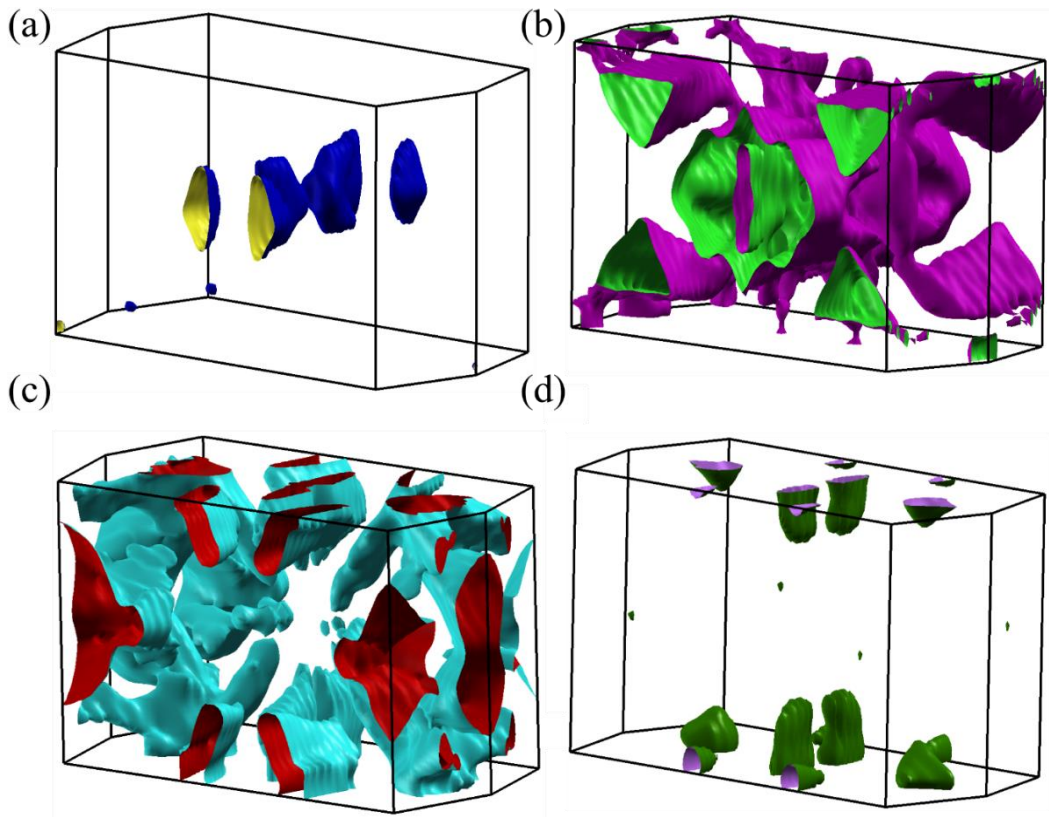

**Supplementary Figure 6:** Fermi surface topology of SrSn4 without SOC that is composed of four bands shown in (a-d).